\RequirePackage{fix-cm}

\documentclass[smallextended]{svjour3}       

\smartqed  

\usepackage{graphicx}
\usepackage{epstopdf}

\usepackage{amsmath}
\usepackage{mathrsfs}

\usepackage{algorithm}
\usepackage{algorithmic}

\usepackage[utf8]{inputenc}
\usepackage[T1]{fontenc}
\usepackage{placeins}
\usepackage{enumitem}
\newlist{steps}{enumerate}{1}
\setlist[steps, 1]{label = Step \arabic*:, leftmargin=3em}

\begin{document}


%

%

\title{Modeling Multiscale Variable Renewable Energy and Inflow Scenarios in Very Large Regions with Nonparametric Bayesian Networks}
\titlerunning{Modeling Multiscale Variable Renewable Energy...}        

\date{March 9, 2020}

\institute{Julio Alberto Dias, Guilherme Machado, Alessandro Soares, Joaquim Dias Garcia \at Praia de Botafogo, 370 - Botafogo, Rio de Janeiro - RJ, 22250-040, Brazil \\
              Tel.: +55-21-3906-2100\\
              \email{\{guilherme, alessandro, joaquim, alberto\}@psr-inc.com}
}
\author{Julio Alberto Dias, Guilherme Machado, Alessandro Soares, Joaquim Dias Garcia}

\maketitle

\begin{abstract}
In this paper, we propose a non-parametric Bayesian network method to generate synthetic scenarios of hourly generation for variable renewable energy(VRE) plants. The methodology consists of a non-parametric estimation of the probability distribution of VRE generation, followed by an inverse probability integral transform, in order to obtain normally distributed variables of VRE generation. Then, we build a Bayesian network based on the evaluation of the spatial correlation between variables (VRE generation and hydro inflows, but load forecast, temperature, and other types of random variables could also be used with the proposed framework), to generate future synthetic scenarios while keeping the historical spatial correlation structure. Finally, we present a real-life case study, that uses real data from the Brazilian power system, to show the improvements that the present methodology allows for real-life studies.
\end{abstract}


\section{Introduction}

In the past, only a relatively small number of countries or regions with a large hydro capacity were concerned with the stochastic modeling of inflows in their planning and operation studies, such as Canada, the US Pacific Northwest, Brazil, Norway, the Mekong region, New Zealand and others. Modeling time dependencies of inflows is complex because of the mixture of long and short-term time effects (e.g. the “flying rivers” that transfer moisture from the Amazon to the central regions of Brazil combined with local rains), cumulative effects (snow-melt), macro climate (El Niño) and others. In turn, the challenges of modeling spatial dependencies of inflows include nonlinearities (combination of surface runoff and aquifer discharge) and the “curse of dimensionality”, related to the exponential increase of cross correlation factors with the number of hydro stations. In Brazil, for example, it is necessary to model about two hundred hydro stations spread out over an area like the continental USA or the EU.

The worldwide insertion of variable renewable energy (VRE) sources such as wind and solar has compounded the above modeling challenges for the following reasons: (i) whereas inflows are usually modeled on a weekly or monthly scale, VRE scenarios require hourly resolution, or shorter \cite{nicolosi2010importance,deane2014impact}; (ii) the “curse of dimensionality” becomes even more severe due to the spatial correlation among hundreds of wind stations and because there may be strong positive and negative correlations between wind and hydro inflows (in Chile, for example, these wind/inflow correlations range from +0.60 to -0.55); (iii) in contrast with inflows, where a lognormal is usually adequate to represent the marginal distribution \cite{contreras2003arima,franses1994model,bezerra2012assessment,pereira1984stochastic}, there is no “family” of distributions for wind, although the Weibull distribution is used for wind speed under very strong assumptions \cite{Tuller1984,seguro2000modern,tuller1984characteristics,dorvlo2002estimating} and Gaussian and Beta distributions are usually used for wind power production forecast \cite{Pinson2012,Bludszuweit2008}.

The importance of developing complex models to generate scenarios that solve the above challenges is that VRE's pose a disruptive change for power system operators, since their intrinsic nature is to have high variability \cite{Holttinen2007}. This impacts the reliability of the system \cite{Doherty2005}, and the best way to access the impact of the VRE's large scale insertion is through the use of simulation tools that has scenarios as its main input.



There is a wide variety of estimation methods in the literature that try to capture the VRE characteristics. The most common approaches are parametric, which requires that the random variables belong to a given family of probability distributions.

The main advantage of this approach is that the distribution depends only on a small set of parameters, allowing an estimation with low computational cost \cite{Zhang2014}. The main issue is when there is not a distribution family that correctly reproduce the behaviour of the historical data. The non-parametric approach does not assumes any kind of shape for the distribution, it will infer the distribution function from the available data. This gives significantly more freedom for the model and allows a better fit for stochastic processes that are not correctly characterized by known distributions. The work in \cite{dias2014non} uses a Kernel Density Estimation (KDE) \cite{Scott1917} to model hydro inflows, and conclude that, since real historical data never fits a exactly log-normal distribution, the non-parametric models perform better. The downside of non-parametric approach is the computational cost which may be significantly greater. Among the methods of non-parametric estimation we highlight: Quantile Regression, Kernel Density Estimation, Probabilistic forecasting based on ensemble and Artificial intelligence \cite{Zhang2014}. In this work we chose the Kernel Density Estimation non-parametric method to model VRE generation, based on \cite{dias2014non}, due to the fact that it is not necessary to assume the distribution family and thus the methodology becomes generic allowing to use the same estimation method for both wind and solar generation, as well as providing a better fit for the wind generation estimation.

Another common problem is the increase in the number of represented variables tends to cause the number of model parameters to grow exponentially. For instance, a multivariate auto-regressive model must consider the cross-correlations between the variables; that is, it must represent the dependence of each variable at a given moment in time in relation not only to itself but also to the other variables at previous instants. This characteristic is usually called the “curse of dimensionality”, referring to the fact that the complexity of a problem increases exponentially in relation to its dimensional size. A common approach for dealing with high dimensional limited data is to rely on regularization techniques such as the $l^1$-regularization used in \cite{souto2014high}.

In this work, it will be presented a methodology that applies a non-parametric approach, based in the KDE estimator, and a Bayesian network to generate synthetic VRE scenarios spatially correlated to hydro inflows. The advantages of this framework are the possibility of considering VRE sources from different types, e.g. wind and solar, using the same methodology with precise representation of each individual marginal distribution. More specifically, the main contributions of this paper are twofold:
\begin{itemize}
    \item  A non-parametric transformation into normally distributed random variables is proposed to transform the original probability distribution of VRE sources generations into a normal distribution. In such approach, Gaussian models, such as the Bayesian network, may be applied in VRE generation transformed data and then transformed back into the original distribution. This idea makes it possible to use any methodology that uses the assumption of normality in any non-gaussian random variable such as VRE generation.
    \item An integrated framework based on a Bayesian network is proposed to capture the spatial correlation between VRE sources and any other random variable. In this approach, the network automatically detects which of the random variables have the most significant correlation with the VRE, so that it is worth to add the random variable to the VRE generation model (avoiding over-fitting). The non-parametric transformation is used to allow the use of the Bayesian network with the assumption of data normality.
\end{itemize}


The paper is organized as follows: in Section \ref{section2}, it will be introduced the general method for estimation using KDE and the Bayesian network; in Section \ref{section3} the procedure defined in \ref{section2} will be applied to generate renewable and inflow future synthetic scenarios, in  Section \ref{casestudy} a case study of the Brazilian system will be presented. Finally, in Section \ref{conclusion} we highlight our conclusions.

\section{Non-parametric Bayesian Model}\label{section2}

In this section we outline the proposed estimation methodology. We start by detailing how non-parametric distributions are used. After that we describe the Bayesian network employed to capture correlations between random variables. Finally, we describe the complete estimation algorithm.

\subsection{Non-parametric distribution}\label{section2.1}
In this work, we are considering that the renewable sources could be of any type, this makes it impossible to assume any distribution family. Therefore, we chose to apply the KDE method to approximate the marginal PDF of each random variable from the historical data.

Once the marginal PDF is obtained, we apply a \textit{Nataf transformation} \cite{Liu1986} to convert the general marginal distribution into a standard normal marginal distribution. This transformation is equivalent to implicitly choosing a Gaussian copula for the random variables \cite{Lebrun2009}. 
In this transform, the vector of random variables $X = (X_1, \dots, X_n)$ will be transformed into a vector of normally distributed random variables $Z = (Z_1, \dots, Z_n)$ for which the correlation can be easily obtained. This model was first presented in \cite{Liu1986} where it is called \textit{Nataf Model} and is also used in the context of wind estimation in \cite{Morales2010}.

This transformation is divided in two steps: first, apply the cumulative probability function to the data to obtain a uniform distribution; second, an normal quantile function to obtain a normal random variable. Because the cumulative probability function $F_{x_i}$ is monotonically increasing and has a one-to-one correspondence, there is an inverse function of the cumulative probability function $X_i = F_{x_i}^{-1}(u_i)$ for each random variable. A new set of historical time series $u_1, u_2, ..., u_{n}$ can be obtained with uniform probability density by applying the previous transformation: 
\begin{align*}
    \left\{\begin{array}{ll}
            u_1 &= F_{x_1}(X_1)\\
            u_2 &= F_{x_2}(X_2)\\
            &\vdots\\
            u_n &= F_{x_n}(X_n)
        \end{array}\right.\\
\end{align*}
At last, we apply the normal quantile function to the uniformly distributed variables, defined formally by:
\begin{equation*}
    Q_N(u_i)=\inf \{z_i\in R:u_i\leq F_{x_i}(X_i)\}
\end{equation*}
This process produces transformed variables that follow a normal distribution.   
\begin{align*}
    \left\{\begin{matrix}z_1=Q_N\left(u_1\right)\\z_2=Q_N\left(u_2\right)\\\begin{matrix}\vdots\\z_{n}=Q_N\left(u_{n}\right)\\\end{matrix}\\\end{matrix}\right.
\end{align*}
By using this transformation, we are assuming that the joint distribution is a multivariate Gaussian distribution and it is possible to calculate the covariance matrix and to estimate the correlation between the random variables.

\subsection{Bayesian network}\label{section2.2}
A Bayesian network \cite{Cooper1992} representation of the statistical dependence between the variables is used to avoid the exponential growth of the model’s complexity as the number of variables increases. Bayesian networks are models that can efficiently and compactly represent the joint probability distribution of n-dimensional variables, by choosing to represent only the most important correlations between variables.
They are characterized by two components:
\begin{itemize}
    \item A directed acyclic graph structure $G=\langle N,E \rangle$, where nodes represents the random variables, $Z =\{Z_1, Z_2, \dots, Z_n\}$, and arcs describe the dependencies and conditional independence between these variables, $E \subseteq N \times N$. We define $pa(Z_i)$ as the parent nodes set of $Z_i$ in $G$. The fundamental property of the Bayesian network is that a node  $Z_i$ is conditionally independent of any other node $Z_j \notin pa(Z_i)$, given $pa(Z_i)$: 
\end{itemize}
    \begin{align*}
        Z_i \perp Z_j | pa(Z_i) \ \forall Z_j \notin pa(Z_i)
    \end{align*}
\begin{itemize}
    \item A set of parameters associated to the graph arcs, that describe the conditional distribution probabilities. From these conditional probability distributions it is possible to reconstruct the joint probability distribution of the random variables:
\end{itemize}
    \begin{equation*}
        P(Z_1, \dots, Z_n) = \prod_{i=1}^n P(Z_i | pa(Z_i))
    \end{equation*}

\begin{figure}
    \centering
    \includegraphics[width=1\linewidth]{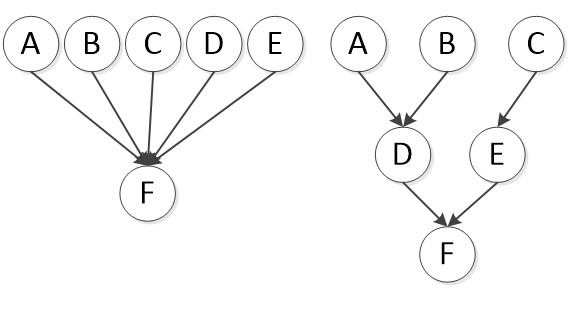}
    \caption{Bayesian network}
    \label{fig:Bayesiannetwork}
\end{figure}

An example of Bayesian network structure is presented in Figure \ref{fig:Bayesiannetwork}, where the left graph is how the correlation is in reality and the graph on the right is how the Bayesian network approximately represents the dependence structure.
The probability distribution $F$ should take into consideration the joint probability distribution of $A, B , C, D$ and $E$ being given:
\begin{equation}
    P(F) = f(P(A, B , C, D, E))
\end{equation}
But assuming that the network structure shown on the right side of the figure is adequate to represent the statistical dependencies of the variables, this function could be written as:
\begin{equation}
    P(F) = f(P(D| A, B) \cdot P(A) \cdot P(B) \cdot P(E|C) \cdot P(C))
\end{equation}
While the exact representation consists of a single probability distribution function, which  has a term that depends on five random variables, in the Bayesian network representation the most complex function is conditioned only to two random variables, $P(D|A, B)$.

The process to estimate the structure of the Bayesian network that best represents the joint probability distribution of a set of random variables has NP-Complete complexity \cite{Junger85,Margaritis2000}, because it is a combinatorial problem in which, in principle, all possible combinations of connections between the nodes should be tested in order to obtain the optimum graph structure. In practice this evaluation is computationally infeasible even for a moderate number of variables. Hence, heuristic methods are applied to reduce the search space and estimate the graph structure, which theoretically is not the optimal structure, but is suitable for the desired representation.

\subsection{Estimation Algorithm}

The required steps for the model estimation of the non-parametric time series are as follows:
\begin{steps}
\item \textbf{Non-parametric estimation}

For each random variable, i.e. each renewable site, fit the KDE from the historical data obtaining one distribution function for every random variable.

\item \textbf{Nataf Transformation}

Transforms the non-parametric marginal distributions into series of normally distributed random variables.

\item \textbf{Bayesian network to determine the statistical dependence}

At this point the Bayesian network is applied on the transformed variables to estimate the statistical dependence. This process creates a directed acyclic graph that maps the statistical dependence structure, generating for each random variable, $q_i$, the set of $pa(q_i)$, in which the Markov condition is observed:
\begin{align*}
    q_i\bot q_j\left|pa\left(q_i\right)\ \forall q_j\right.\notin\ pa\left(q_i\right)
\end{align*}

\item \textbf{Estimation of the regression parameters}

The regression model of each random variable, $X_i$,  is given by:
\begin{align*}
    X_i=F_{x_i}^{-1}\bigg(F_N\bigg(\sum_{q_j\in p a(q_i)} q_ja_{ij}+\varepsilon_i\bigg)\bigg)
\end{align*}
Where, $pa(n_i)$ is the set of parent nodes of $q_i$ in the statistical dependency structure described by the Bayesian network;
$a_{ij}$ are the regression parameters that relate $q_i$  to its parents: $q_j\in pa(q_i)$;
$\varepsilon_i$ is the random variable describing the innovation of $X_i$;
$F_N$ is the normal cumulative distribution function, corresponding to the inverse of the $Q_N$ function.
For each $X_i$, the best parameters that fit the historical dataset are obtained through a least square estimation approach.
\end{steps}

\section{VRE and inflow scenarios generation}\label{section3}
Once the model is estimated, the required steps for the simulation are described in the next sections:

\subsection{Multivariate sampling of the normal variables}

In this context, multivariate sampling of the normal variables consists of obtaining normally distributed discrete samples using the Bayesian network structure information and the adjusted parameters. A sample $s$ of the set of random variables $q^s=\left[q_1^s,q_2^s,\dots,q_{n}^s\right]$ described by the Bayesian network, is obtained by recursively following the graph nodes of the Bayesian network from the parent nodes to the child nodes.
For each random variable, we sample the innovation random variable $\epsilon_i^s$ from its empirical distribution, and calculate $q_i^s$ applying the sampled values of its parent nodes in the fitted expression: 
\begin{align}
    {q_i}^s=\sum_{q_j\in p a\left(q_i\right)}{q_j}^s a_{ij}+{\varepsilon_i}^s
\end{align}

\subsection{Transformation of the normal variables into the uniform distribution}

The uniform random variable samples are obtained by applying the normal cumulative probability function $F_N$ to each variable ${q_i}^s$ according to:
\begin{align}
    \left\{\begin{matrix}u_1^S=F_N\left(q_1^S\right)\\u_2^S=F_N\left(q_2^S\right)\\\begin{matrix}\vdots\\u_n^S=F_N\left(q_n^S\right)\\\end{matrix}\\\end{matrix}\right.
\end{align}

\subsection{Transformation of the uniform variables into the original distribution}

This process is used to obtain samples with the same distribution as the historical series. The uniform random variable samples are transformed by applying the inverse transformation function, $F_{x_i}^{-1}$, to each variable, $u_i^s$, according to:
\begin{align}
    \left\{\begin{matrix}x_1^s=F_{x_1}^{-1}\left(u_1^s\right)\\x_2^s=F_{x_2}^{-1}\left(u_2^s\right)\\\begin{matrix}\vdots\\x_{n}^s=F_{x_n}^{-1}\left(u_{n}^s\right)\\\end{matrix}\\\end{matrix}\right.
\end{align}

\subsection{Fitting hydro and VRE correlation with different resolution }\label{disagregating}

In the estimation phase, depending on the hydrological model chosen, it might happen that the hydro generated scenarios have a monthly or weekly resolution and the renewable historical time series has an hourly resolution. To be able to capture the variability of the VRE sources, the generated scenarios must have at least an lower resolution, e.g. an hourly resolution. In this situation it is also possible to generate hourly synthetic VRE scenarios, following the procedure that will be described next.

The nonparametric Bayesian model will be fitted using the same resolution as the generated inflows, and the VRE scenarios will latter be disaggregated into an hourly resolution. For that, the historical scenarios of the VRE must first be aggregated to fit the nonparametric Bayesian model. After, a pos-processing is needed. This pos-processing consists of first applying a Principal Components Analysis (PCA) decomposition in each month/week of the VRE aggregated historical data. This will result in a decomposition matrix for each month/week.

To transform the generated monthly/weekly scenarios into hourly scenarios, the decomposition matrix will be applied in both the aggregated historical data and the generated scenarios. For each transformed generated scenario, we calculate the squared  euclidean distance to each transformed historical scenario and select the historical time series that has the lowest distance as a profile for this generated scenario. Finally, we disaggregate the montly/weekly scenario into the historical hourly profile, adjusting the average capacity factor of the VRE to match the one of the generated scenario.


\section{Case Study}

The presented methodology will be applied to the Brazilian electrical system, which contains 180 hydro stations and 87 VRE plants, between them 38 are wind power plants, 10 are concentrated solar power (CSP), 28 are distributed generation solar plants (DGSP) and 5 are small hydro stations. 

The Brazilian system makes a good case study for the presented methodology due to the number of hydro stations and VRE plants. If a standard model (estimating the correlation matrix of all random variables) was used to fit all the renewable stations, the number of parameters of this model would be 378006 for 105420 observations. In this case, the model would have 3.59 times more parameters than data. In the presented approach, thanks to the Bayesian Network the estimated model has only 9441 parameters, which is equivalent to 9\% of the number of observations and 2.5\% of the number of parameters of the standard approach. While the number of parameters was greatly reduced, the quality of the estimation it is not, as the Bayesian network captures the most relevant correlations. This shows the strength of presented methodology.

As described in the Section \ref{section3}, the hydro inflow scenarios are generated before the VRE scenarios. The statistical model chosen to generate hydro inflow scenarios is a log-normal AR(p) one \cite{Matalas1967,Hipel1994}. The VRE scenarios are generated with the KDE method having hourly resolution, while the inflow scenarios have monthly resolution. The model first aggregates the capacity factor of the hourly historical time series into a monthly time series. Secondly, it fits the nonparametric Bayesian model and finally it generates the synthetic scenarios, that are latter disaggregated into an hourly time series following \ref{disagregating}. The fact that the renewable scenarios do not need to have the same granularity as the hydro inflow allows the generated VRE scenarios to capture the variability of these sources in a greater level of detail.

For this analysis, it was generated 100 scenarios for each station, from a historical database of 34 years. Three VRE stations were chosen randomly to display the results: one wind power plant, one CSP plant and one DGSP.

\begin{figure}[htb]
  \centering
  \includegraphics[width=1\linewidth]{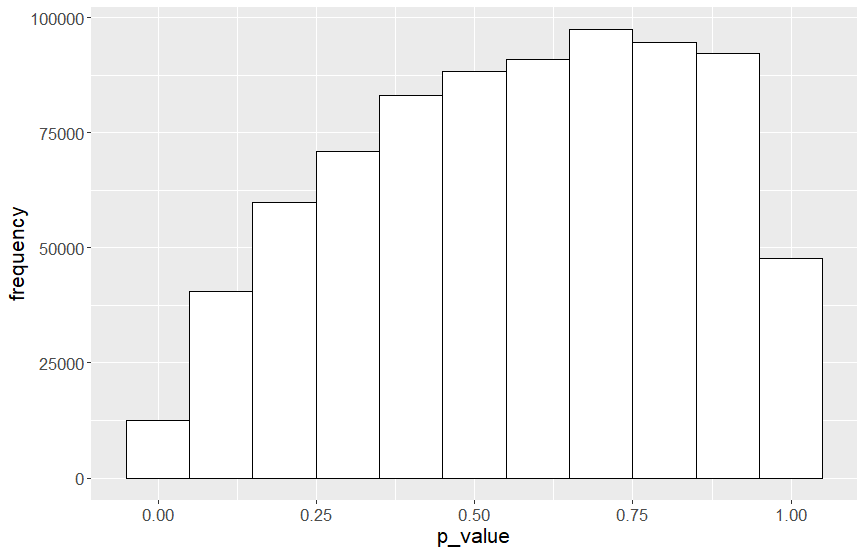}
  \caption{Fischer's Z test}
    \label{fig:fischer}
\end{figure}

The generated scenarios were validated with a Fischer's Z \cite{fischer1925} test and it attend the null hypothesis that the 96\% of the correlations are the same for a significance level of 10\%. The histogram of the statistical test is displayed in the Figure \ref{fig:fischer}.

\begin{figure}[htb]
  \centering
  \includegraphics[width=1\linewidth]{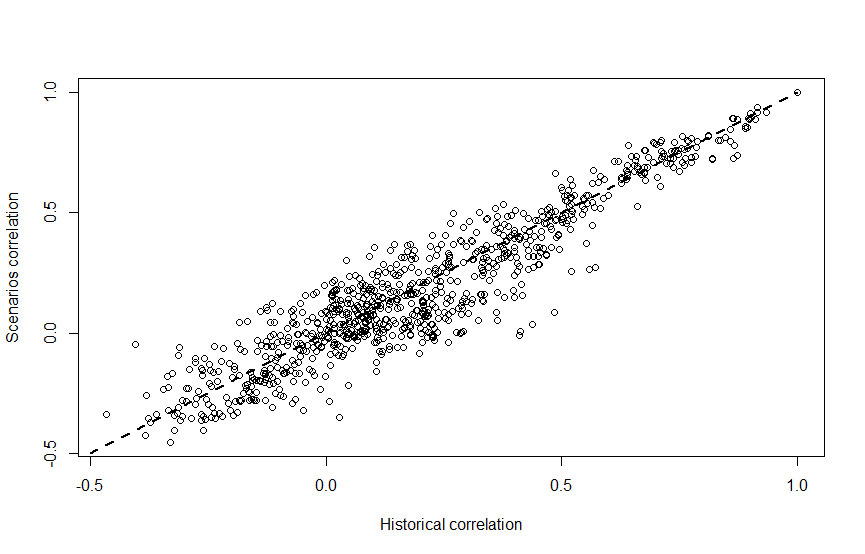}
  \caption{Synthetic vs Historical correlation}
    \label{fig:synthist}
\end{figure}

The synthetic and the historical correlations between all renewables are displayed in a scatter plot in the Figure \ref{fig:synthist}. It shows that the synthetic scenarios can reproduce well the historical correlation, as the scatter plot is concentrated around the identity line, i.e. the line of equality.

\begin{figure}[ht]
  \centering
  \includegraphics[width=1\linewidth]{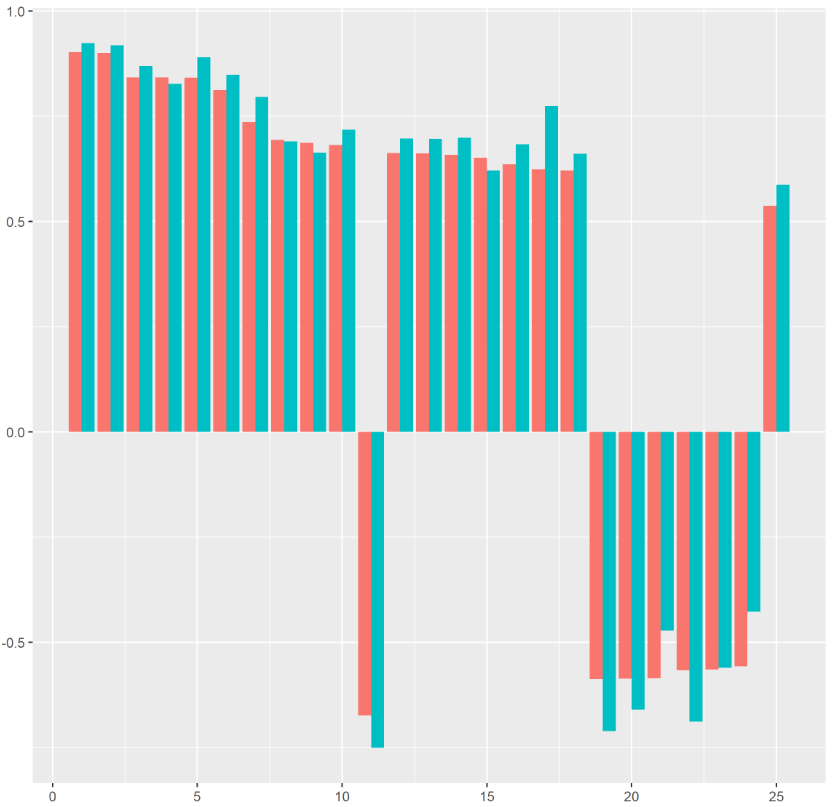}
  \caption{DGSP correlation to other VRE stations}\label{cor:gd}
\end{figure}

One last check for the correlation of the generated scenarios is displayed in Figure \ref{cor:gd}, where the DGSP correlation to the other VRE stations are presented for both the historical and synthetic time series. The historical correlations are displayed in red, while the synthetic correlations are displayed in green. The y-axis contains the value of the correlations and the x-axis contains the other renewable stations. In this plot, only the stations with the highest correlation to the DGSP are shown, i.e. the most significant ones.

\begin{figure}[ht]
  \centering
  \includegraphics[width=1\linewidth]{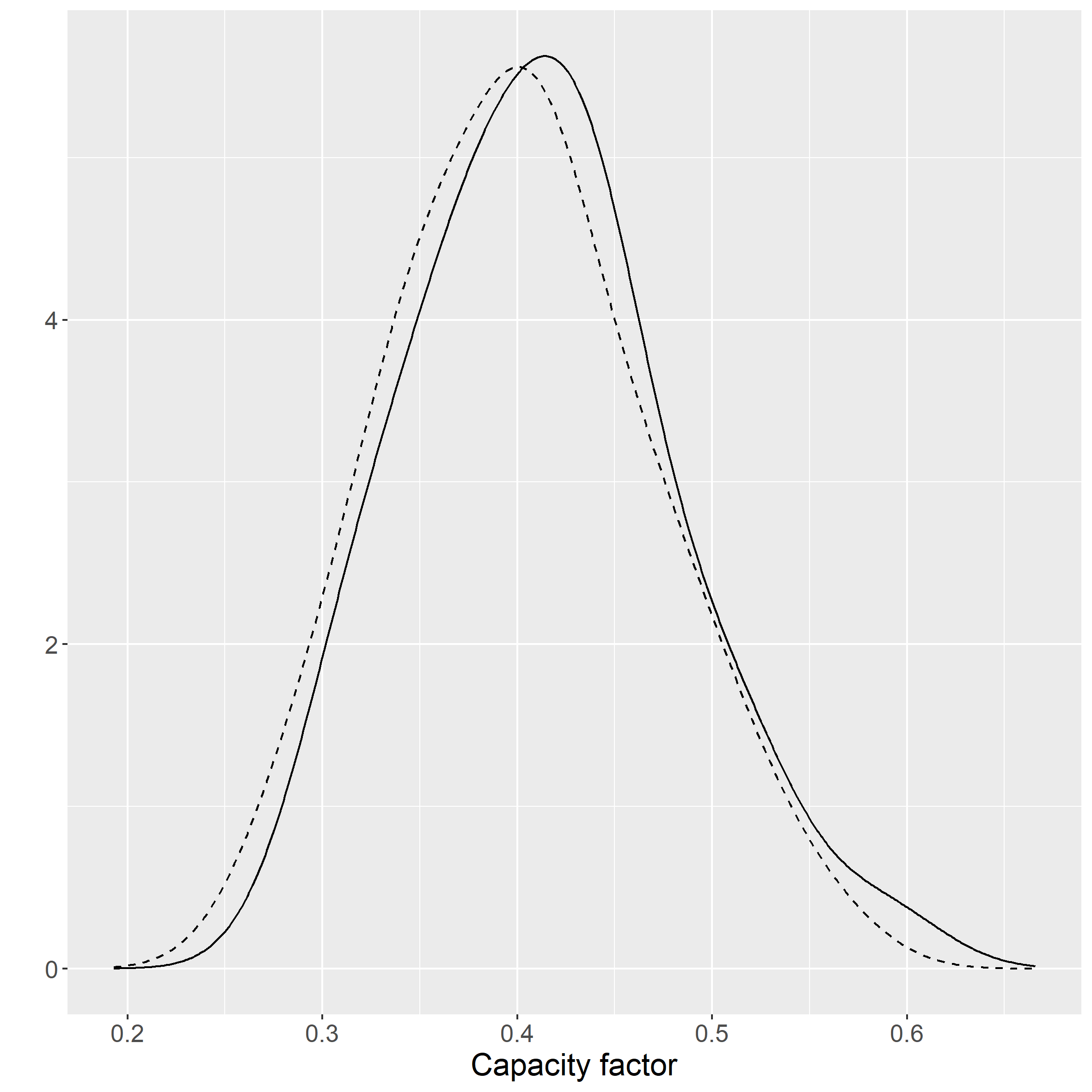}
  \caption{Wind station PDF}\label{pdf:eol}
\end{figure}

The PDF of the synthetic scenarios and also of the historical data set for the wind station is displayed at the Figure \ref{pdf:eol}. The dashed line is the synthetic generated PDF, while the solid line is the historical PDF. This figure shows that the PDF of the historical data is well reproduced by the synthetic scenarios, mimicking the overall behaviour of the random variable. 

\begin{figure}[ht]
  \centering
  \includegraphics[width=1\linewidth]{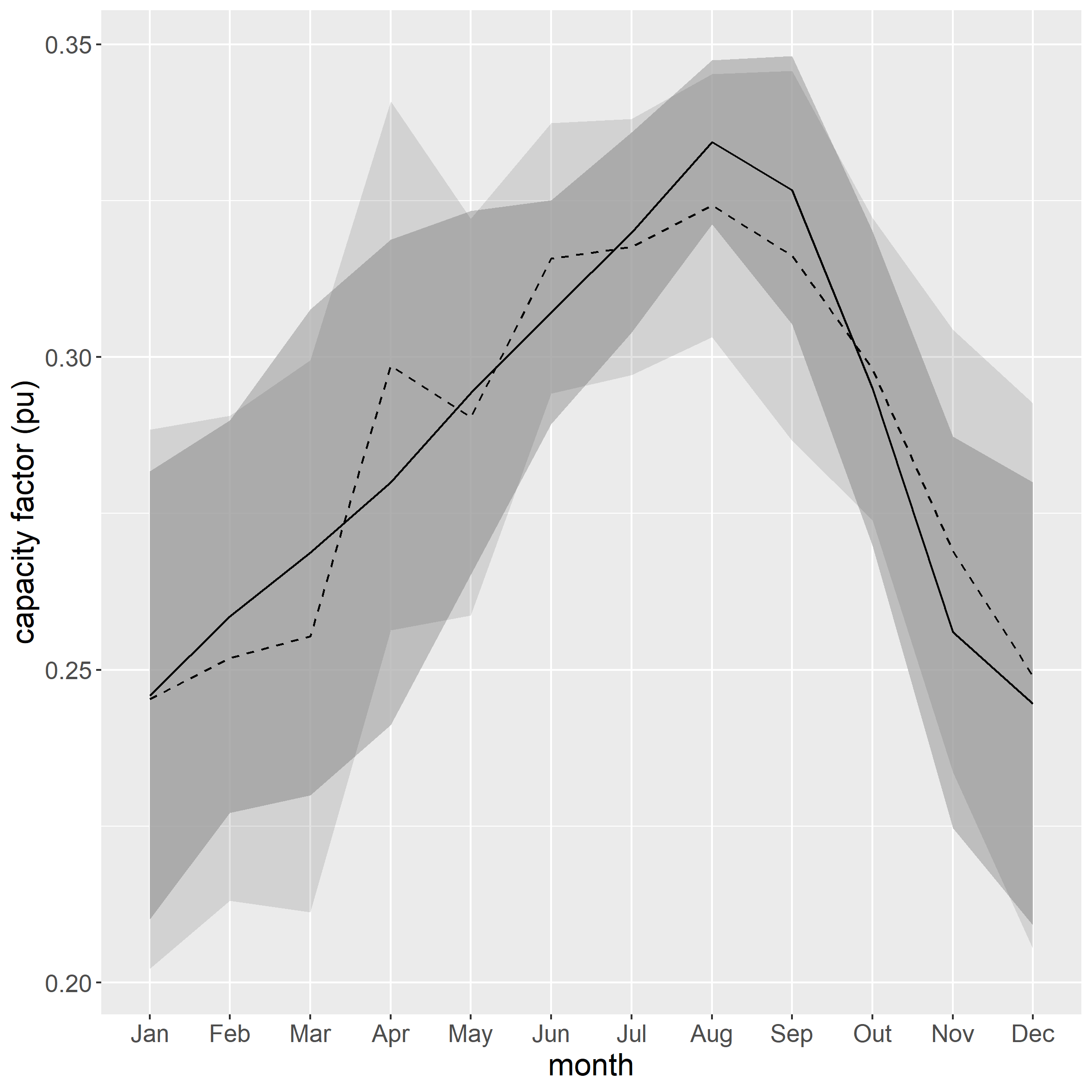}
  \caption{Confidence interval for CSP station}\label{conf:csp}
\end{figure}

Looking more closely to the generated scenarios, the Figure \ref{conf:csp} displays the confidence interval around the average for historical and synthetic scenarios in the time-frame of one year for the CSP station. Again, the dashed curve is the synthetic scenario and the solid is the historical one. In light grey, we display the confidence interval of the synthetic scenarios and, in dark grey, the confidence interval of the historical scenario. Not only is the synthetic average inside the confidence interval of the historical data but it is also close to the historical average.

\begin{figure}[ht]
  \centering
  \includegraphics[width=1\linewidth]{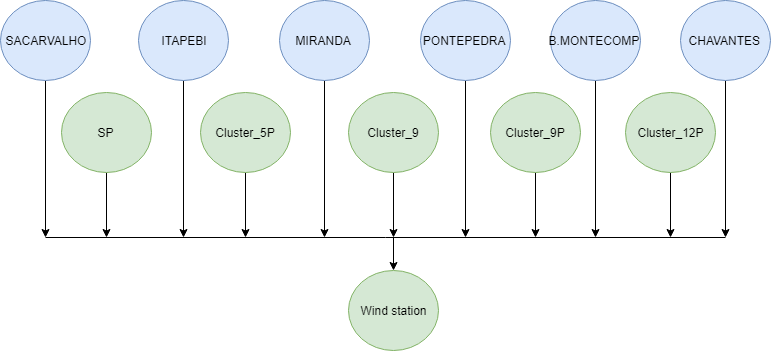}
  \caption{Wind station Bayesian network}\label{fig:bayesiannetwork}
\end{figure}

The most significant correlations chosen by the Bayesian network for the wind station is displayed in the Figure \ref{fig:bayesiannetwork}, where the wind station is correlated to 6 hydro stations and 5 VRE stations. The VRE stations are the green colored circles and the hydro stations are the blue circles.

\FloatBarrier\label{casestudy}

\section{Conclusion}\label{conclusion}

In this paper, we proposed a methodology to generate future synthetic VRE scenarios that may represent the spatial correlation with hydro inflows or any other random variable. The idea behind the methodology is to use the Bayesian network to search for random variables with a significant correlation with VRE generation so that it can be used in order to generate the synthetic scenarios.

The non-parametric transformation applied to the VRE generation is used to calculate a function that transforms the original data into a normally distributed data, which makes it possible to use the Bayesian network with the assumption of data normality. This transformation may also be used for any other data that may have any kind of correlation with the VRE generation, such as demand, temperature, rain, and others.

The methodology used in this paper is only addressing the correlation between VRE generation and hydro inflows. But since the Bayesian network is a generic model, it's possible to add any random variable, correlate it with the VRE generation and automatically decide whether it is worth to consider these variables into the VRE generation model, or not. So, in other words, the Bayesian network acts as a filter, preventing to get models with a huge number of parameters, i.e., avoiding to try to correlate each VRE source with all of the other random variables in the network (hydro plants in this paper).

In the case study presented here, for the Brazilian system, with 180 hydro stations and 87 VRE stations, we tried to capture the spatial dependence between VRE generation and hydro inflows. We showed that the generated scenarios reproduce very well the historical correlations and PDFs, while only using 2.5\% of the number of parameters of a standard approach. The number of parameters is equivalent to 9\% of the observations and the correlations were validated using the Fischer's Z test.

\clearpage


\bibliographystyle{ieeetr}
\bibliography{references}

\clearpage

\end{document}